Aldosterone and Dexamethasone Activate African Lungfish Mineralocorticoid Receptor:

Increased Activation After Removal of the Amino-Terminal Domain

Yoshinao Katsu[1,2], Shin Oana[2], Xiaozhi Lin[2], Susumu Hyodo[3], Michael E. Baker[4]

[1]Graduate School of Life Science

Hokkaido University

Sapporo, Japan

[2]Faculty of Sciences

Hokkaido University

Sapporo, Japan

[3]Laboratory of Physiology

Atmosphere and Ocean Research Institute

University of Tokyo

Chiba, Japan

[4]Division of Nephrology-Hypertension

Department of Medicine, 0693

University of California, San Diego

9500 Gilman Drive

La Jolla, CA 92093-0693

Center for Academic Research and Training in Anthropogeny (CARTA)

University of California, San Diego

La Jolla, CA 92093

Correspondence to:

Y. Katsu:  ykatsu@sci.hokudai.ac.jp

M. E. Baker: mbaker@ucsd.edu

**ABSTRACT.** Aldosterone, the main physiological mineralocorticoid in humans and other terrestrial vertebrates, first appears in lungfish, which are lobe-finned fish that are forerunners of terrestrial vertebrates. Aldosterone activation of the MR regulates internal homeostasis of water, sodium and potassium, which was critical in the conquest of land by vertebrates. We studied transcriptional activation of the slender African lungfish MR by aldosterone, other corticosteroids and progesterone and find that aldosterone, 11-deoxycorticosterone, 11-deoxycortisol and progesterone have half-maximal responses (EC50s) below 1 nM and are potential physiological mineralocorticoids. In contrast, EC50s for corticosterone and cortisol were 23 nM and 66 nM, respectively. Unexpectedly, truncated lungfish MR, consisting of the DNA-binding, hinge and steroid-binding domains, had a stronger response to corticosteroids and progesterone than full-length lungfish MR, indicating that the N-terminal domain represses steroid activation of lungfish MR, unlike human MR in which the N-terminal domain contains an activation function. BLAST searches of GenBank did not retrieve a GR ortholog, leading us to test dexamethasone and triamcinolone for activation of lungfish MR. At 10 nM, both synthetic glucocorticoids are about 4-fold stronger than 10 nM aldosterone in activating full-length lungfish MR, leading us to propose that lungfish MR also functions as a GR.

## INTRODUCTION

The mineralocorticoid receptor (MR) and glucocorticoid receptor (GR) belong to the nuclear receptor family, a diverse group of transcription factors that arose in multicellular animals [1–3]. The MR and GR have key roles in the physiology of humans and other terrestrial vertebrates and fish [4–11]. The MR and GR evolved from an ancestral corticoid receptor (CR) in a jawless fish (cyclostome), which has descendants in modern lampreys and hagfish [12–14]. A distinct MR and GR first appear in cartilaginous fishes (Chondrichthyes) [1,13,15–17], which diverged from bony vertebrates about 450 million years ago [18,19].

Aldosterone is the main physiological mineralocorticoid in humans and other terrestrial vertebrates [5,6,9,20–23]. Aldosterone activation of the MR in the kidney regulates salt and water homeostasis by promoting sodium and water reabsorption and potassium secretion, a mechanism that conserves salt and water. Thus, it is puzzling that aldosterone is a potent transcriptional activator of lamprey and hagfish CRs [15], skate MR [16] and elephant shark MR [13,24] because aldosterone is not synthesized by lampreys [15], cartilaginous fishes or ray finned fishes [25]. Aldosterone first appears in lungfish [26–28], which are lobe-finned fish that are forerunners of terrestrial vertebrates [29–31]. The key phylogenetic position of lungfish in the transition of vertebrates from water to land [27,29,30,32] and the role of the MR in maintaining internal electrolyte homeostasis [5,8,33,34] motivated us to investigate the response of the slender African lungfish MR to aldosterone, cortisol and other corticosteroids (Figure 1), as well as activation by progestins, which also activate elephant shark MR [24], ray-finned fish MR [35–39] and chicken MR [24,40].

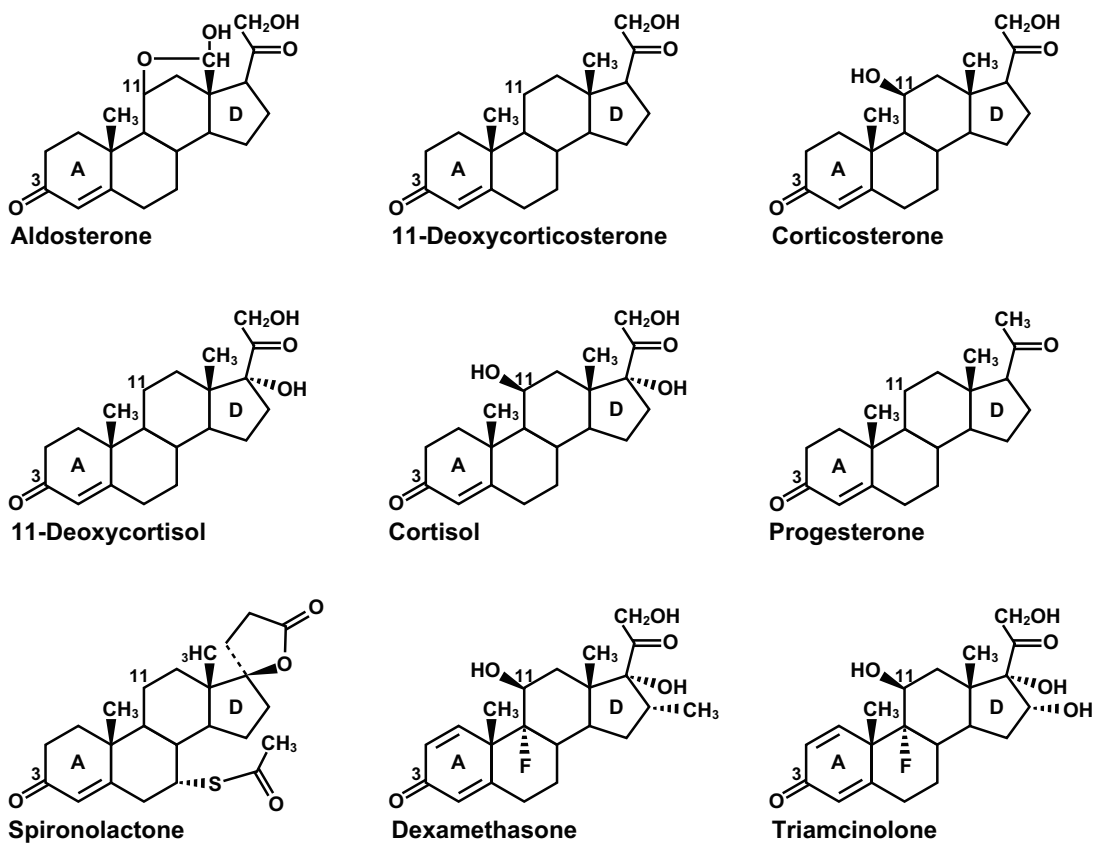

**Figure 1. Structures of Corticosteroids, Dexamethasone, Triamcinolone, Progesterone and Spironolactone.** Aldosterone and 11-deoxycorticosterone are mineralocorticoids [41]. 11-deoxycortisol is a mineralocorticoid in lamprey [42,43]. Cortisol and corticosterone are glucocorticoids in terrestrial vertebrates and ray-finned fish [41,44]. Dexamethasone and triamcinolone are synthetic glucocorticoids. Progesterone is female reproductive steroid that also is important in male physiology [45,46]. Spironolactone is a mineralocorticoid antagonist in humans [47,48].

Our investigation also uncovered an unexpected role of the N-terminal domain (NTD) of lungfish MR in inhibiting transcriptional activation by steroids. Like other steroid receptors, lungfish MR is a multi-domain protein, consisting of an NTD (domains A and B), a central DNA-binding domain (DBD) (domain C), a hinge domain (D) and a C-terminal ligand-binding domain (LBD) (domain E) [49–51] (Figure 2). The NTD in the human MR contains an activation function domain (AF1), which is split into two segments [49–51]. (Figure 2). As described below, we find that in contrast to human MR [24,49,51,52], the NTD in full-length lungfish MR reduces steroid-mediated activation of lungfish MR, compared to truncated lungfish MR-CDE in cells transfected with a 3X-Tyrosine Amino Transferase (TAT3) promoter [53].

We also find that lungfish MR is activated by dexamethasone. At 10 nM, dexamethasone activates full-length lungfish MR and truncated lungfish MR with a signal that is 4-fold and 6-fold stronger, respectively, than that of 10 nM aldosterone. This strong response to dexamethasone and the absence of a lungfish GR sequence after a BLAST search of GenBank leads us to propose that lungfish MR also functions as a GR.

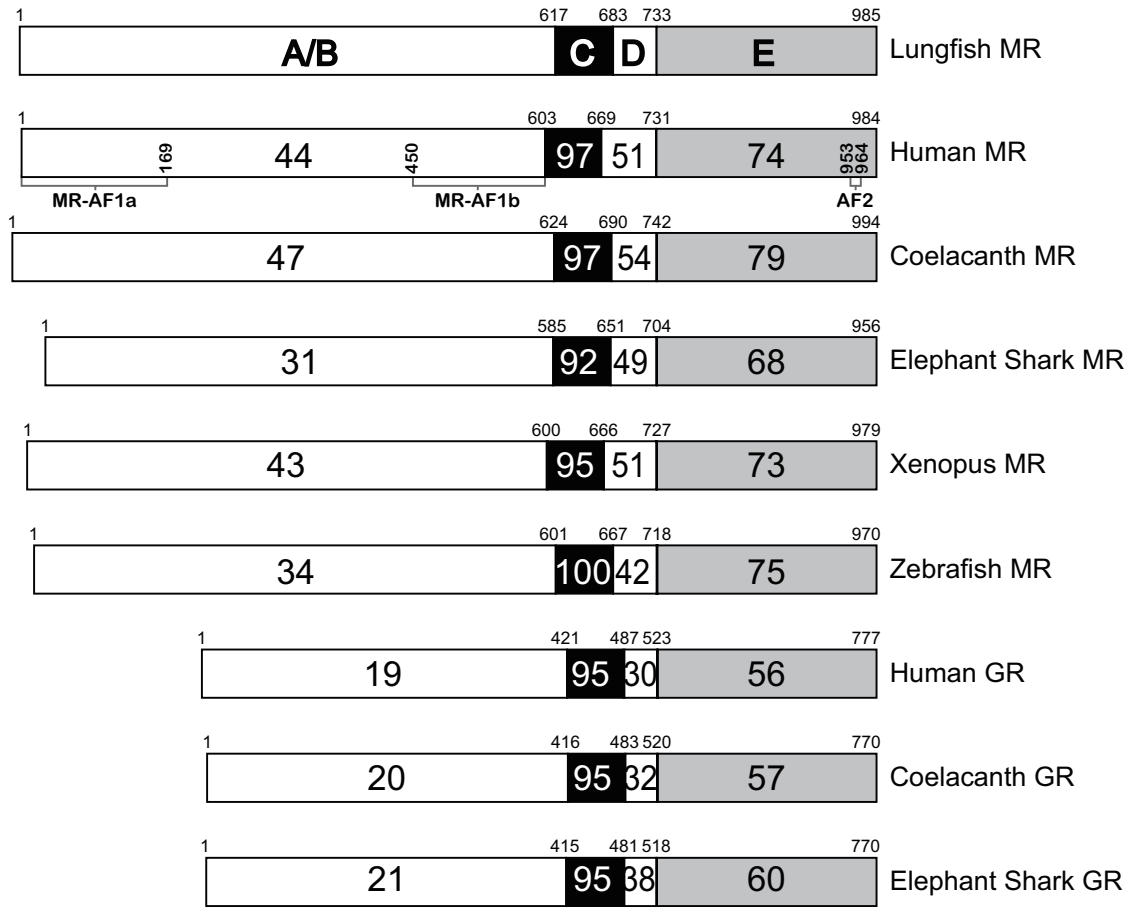

**Figure 2. Comparison of the functional domains of lungfish MR to corresponding domains in selected vertebrate MRs (human, coelacanth, elephant shark, *Xenopus*, zebrafish) and GRs (human, coelacanth, elephant shark).** Lungfish MR and human MR have 97% and 74% identity in DBD and LBD, respectively. Lungfish MR and elephant shark MR have 92% and 68% identity in DBD and LBD, respectively. This strong conservation of the DBD and LBD contrasts with the low sequence identity of 44% and 47% between their NTDs. There are similar % identities between corresponding domains in lungfish MR and other MRs.

## RESULTS

**Transcriptional activation of full-length and truncated lungfish MR by corticosteroids, progestins and dexamethasone.**

We screened a panel of steroids (Figure 1) at 10 nM for transcriptional activation of full-length and truncated lungfish MR containing the CDE domains (MR-CDE) using two promoters: 2X-Mouse Mammary Tumor Virus (MMTV) [54,55] and TAT3 [53], which along with plasmids for both lungfish MRs were transfected into HEK293 cells.

As shown in Figure 3A, there was about 2 to 3-fold activation by 10 nM aldosterone, other corticosteroids or progesterone of full-length lungfish MR using the MMTV-luc reporter and less steroid activation of lungfish MR-CDE (Figure 3B).

Interestingly, compared to activation of full-length lungfish MR with the MMTV promoter (Figure 3A), transcriptional activation of full-length lungfish MR with a TAT3 promoter and 10 nM aldosterone, other physiological corticosteroids or dexamethasone increased by about 1.5 to 2-fold (Figure 3C). Unexpectedly, lungfish MR-CDE with the TAT3 promoter had an additional 2-fold increase in activation by all corticosteroids (Figure 3D). Progesterone activated lungfish MR in accord with the prediction of Fuller et al. [37,39,56]. Together, these experiments show that removal of the NTD increases corticosteroid and progesterone activation of lungfish MR in the presence of the TAT3 promoter.

Our results with dexamethasone, which activates human MR [52,57–59], were unexpected. To our surprise, compared to aldosterone, dexamethasone was about 3-fold and 6-fold more active, respectively, in activating full-length lungfish MR (Figure3C) and truncated lungfish MR (Figure 3D) with the TAT3 promoter. Moreover, both cortisol and corticosterone have stronger fold-activation than does aldosterone of lungfish MR using the TAT3 promoter. Under these conditions, lungfish MR appears to have a GR-like response to steroids.

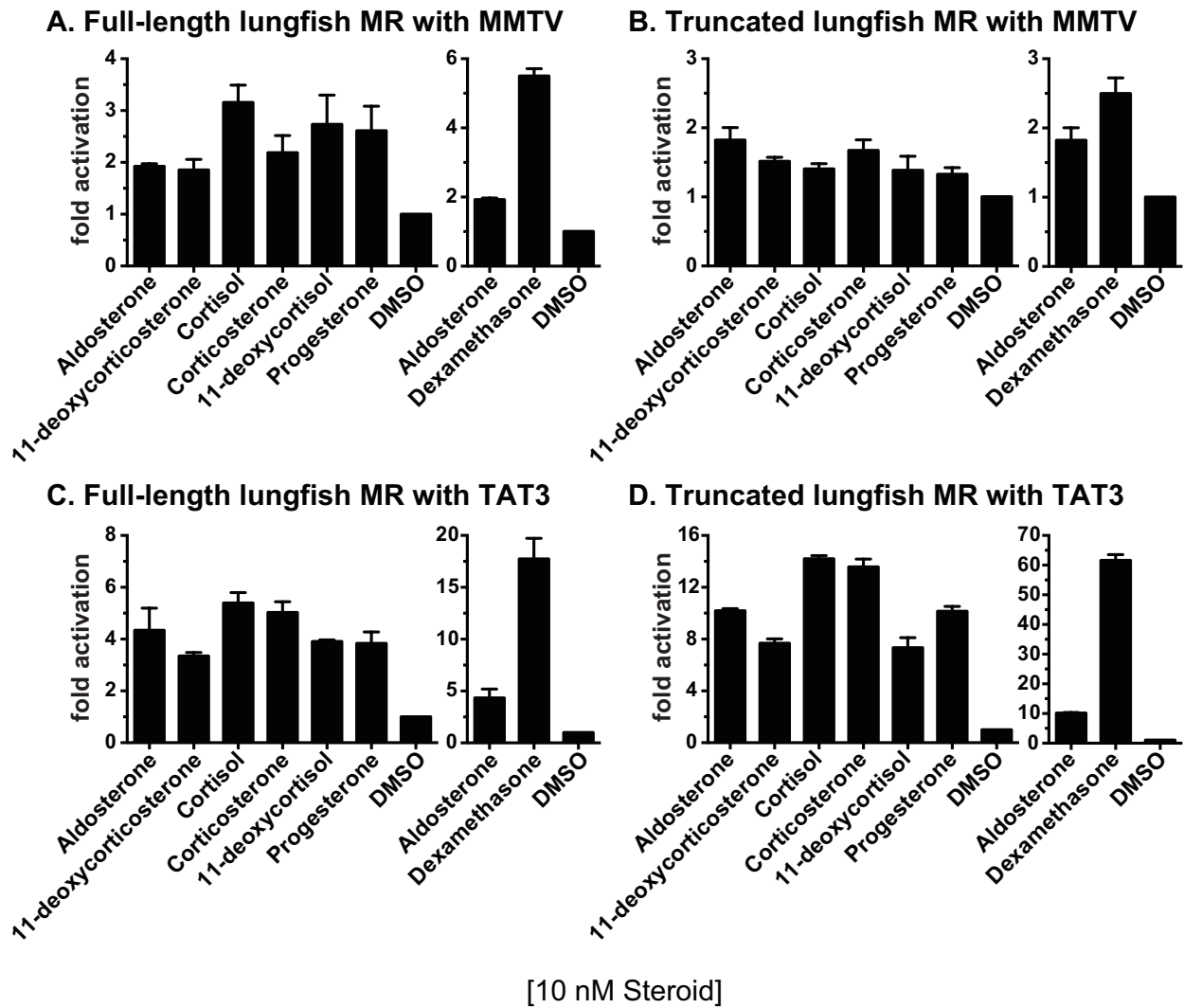

**Figure 3. Ligand specificity of full-length and truncated lungfish MR.**
Plasmids for full-length lungfish MR or truncated lungfish MR (MR-CDE) were expressed in HEK293 cells with an MMTV-luciferase reporter or a TAT3-luciferase reporter. Transfected cells were treated with either 10 nM aldosterone, cortisol, 11-deoxycortisol, corticosterone, 11-deoxycorticosterone, progesterone, dexamethasone or vehicle alone (DMSO). Results are expressed as means ± SEM, n=3. Y-axis indicates fold-activation compared to the activity of control vector with vehicle alone as 1. A. Full-length lungfish MR with MMTV-luciferase. B. Truncated lungfish MR (MR-CDE) with MMTV-luciferase. C. Full-length lungfish MR with TAT3-luciferase. D. Truncated lungfish MR (MR-CDE) with TAT3-luciferase.

**Spironolactone and eplerenone are transcriptional activators of lungfish MR.**

Because spironolactone, an antagonist of human MR, activates elephant shark MR [24], zebrafish MR [37,40,60] and trout MR [38], we investigated spironolactone for activation of full-length lungfish MR and truncated lungfish MR-CDE. We also studied activation by eplerenone, another antagonist of human MR [61]. As shown in Figure 4, both spironolactone and eplerenone activated lungfish MR with a TAT3 promoter, and there was a further increase in fold-activation by both steroids of lungfish MR-CDE.

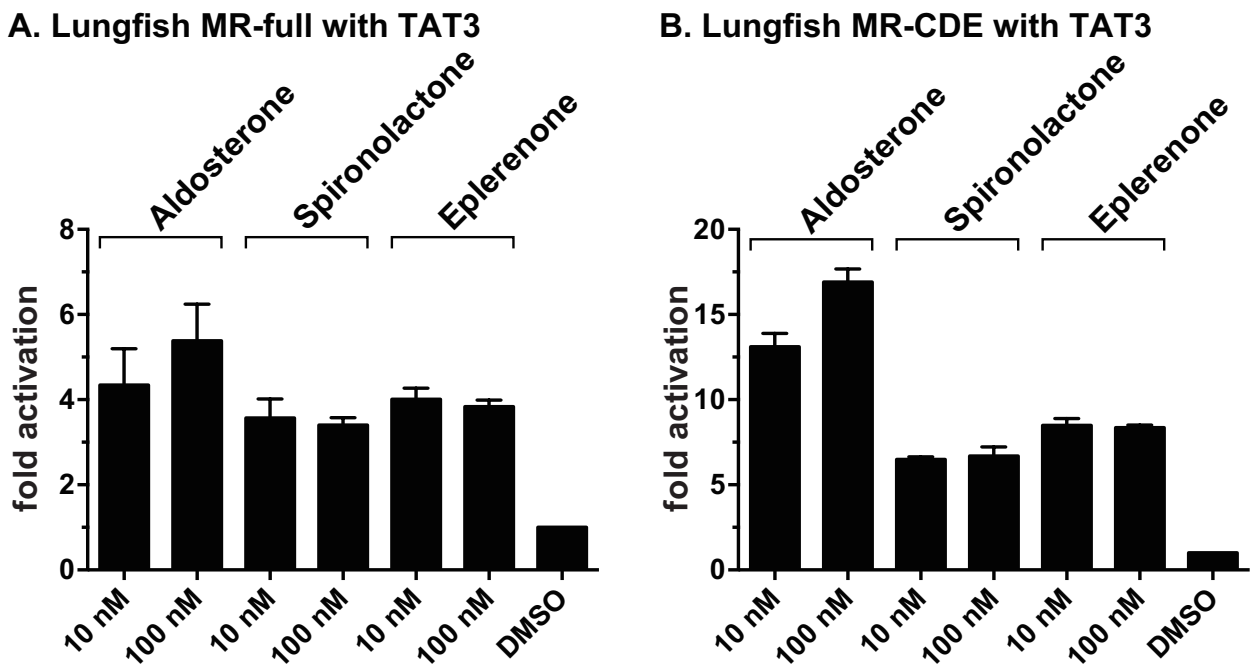

**Figure 4. Spironolactone and eplerenone activation of full-length and truncated lungfish MR.** Plasmids for full-length lungfish MR or truncated lungfish MR (MR-CDE) were expressed in HEK293 cells with a TAT3-luciferase reporter. Transfected cells were treated with either 10 nM or 100 nM aldosterone, spironolactone or eplerenone or vehicle alone (DMSO). Results are expressed as means ± SEM, n=3. Y-axis indicates fold-activation compared to the activity of control vector with vehicle alone as 1. A. Full-length lungfish MR with TAT3. B. Truncated lungfish MR (MR-CDE) with TAT3.

**Concentration-dependent activation by corticosteroids and progestins of full-length and truncated lungfish MR**.

To gain a quantitative measure of corticosteroid and progestin activation of full-length and truncated lungfish MR, we determined the concentration dependence of transcriptional activation by corticosteroids and progestins of full-length lungfish MR and lungfish MR-CDE using TAT3 (Figure 5). This data was used to calculate a half maximal response (EC50) for steroid activation of lungfish MR with a TAT3 promoter (Table 1). For full-length lungfish MR, the four lowest EC50s were for aldosterone (0.04nM), 11-deoxycorticosterone (0.04 nM), 11-deoxycortisol (0.17nM) and progesterone (0.03nM). These low EC50s are consistent with a physiological role for one or more of these steroids as ligand for lungfish MR. In contrast, corticosterone and cortisol, two physiological corticosteroids in terrestrial vertebrates, had EC50s of 23.1nM and 66.1nM, respectively. Two synthetic glucocorticoids, dexamethasone and triamcinolone, had EC50s of 4.7nM and 1.3nM, respectively.

For truncated lungfish MR, there were similar low EC50s for aldosterone (0.24nM), 11-deoxycorticosterone (0.013nM), 11-deoxycortisol (0.27nM) and progesterone (0.04nM). EC50s for corticosterone and cortisol were 85.5nM and 86.7nM, respectively. EC50s for dexamethasone and triamcinolone were 7.7nM and 2.4nM, respectively.

Overall, these results reveal that the EC50s of aldosterone, 11-deoxycorticosterone, 11-deoxycortisol and progesterone for full-length lungfish MR and lungfish MR-CDE are similar and that one or more of these steroids could be a physiological mineralocorticoid in lungfish. Although EC50s for full-length lungfish MR of triamcinolone and dexamethasone were at least 10-fold higher than that of aldosterone, deoxycorticosterone, 11-deoxycortisol and progesterone, compared to these steroids, dexamethasone and triamcinolone have a several fold higher activation of full-length and truncated lungfish MR (Figure 5). Consistent with data in Figure 3, deletion of the NTD to form truncated lungfish MR-CDE increased fold-activation by aldosterone, the other corticosteroids, progesterone, dexamethasone and triamcinolone. However, deletion of the NTD did not have a large effect on their EC50s.

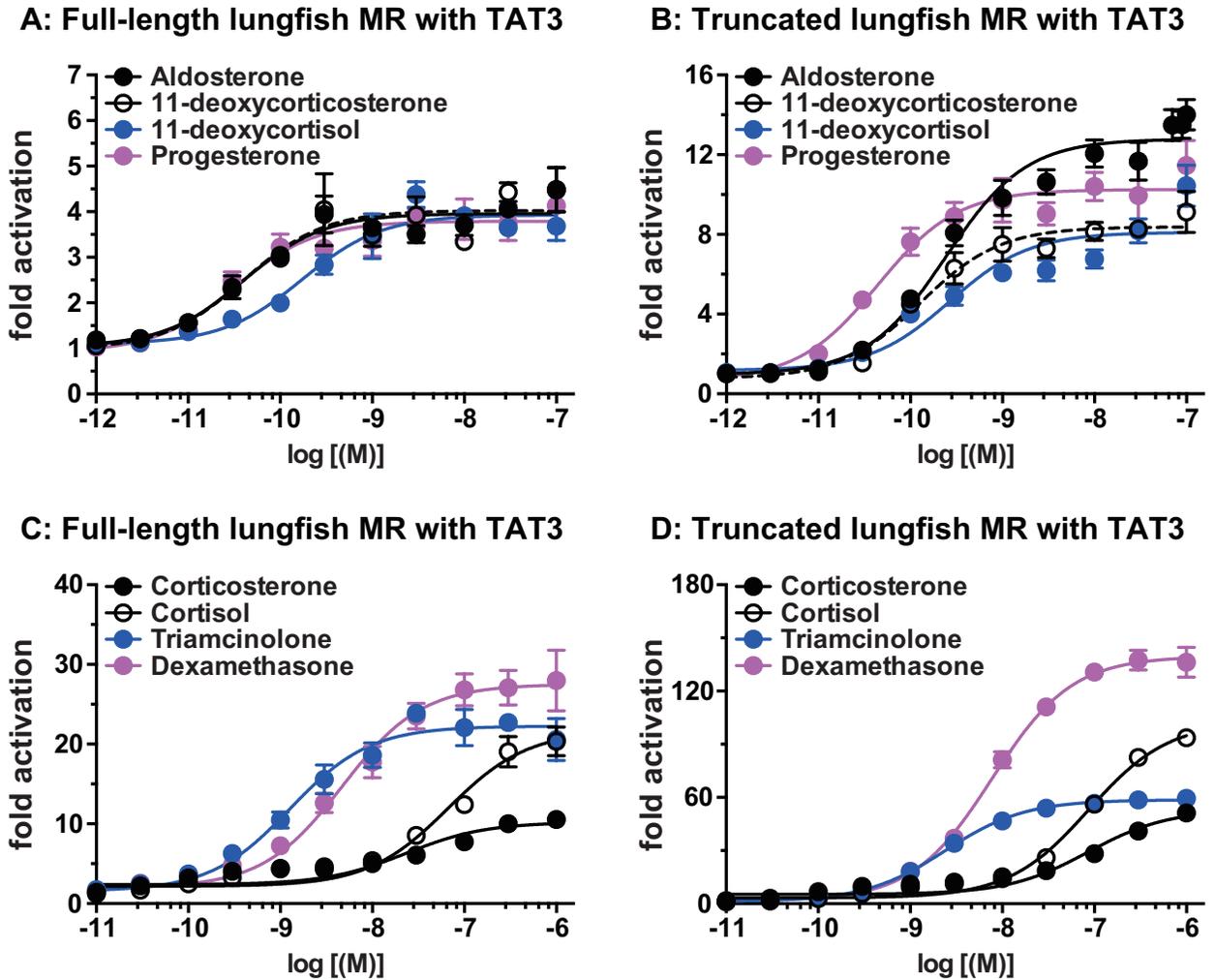

**Fig. 5. Concentration-dependent transcriptional activation by corticosteroids, progesterone, dexamethasone and triamcinolone of full length and truncated lungfish MR.** Plasmids for full-length lungfish MR or truncated lungfish MR, were expressed in HEK293 cells with a TAT3-luciferase promoter. Cells were treated with increasing concentrations of either aldosterone, cortisol, corticosterone, 11-deoxycorticosterone, 11-deoxycortisol, progesterone, dexamethasone and triamcinolone or vehicle alone (DMSO). Results are expressed as means ± SEM, n=3. Y-axis indicates fold-activation compared to the activity of control vector with vehicle (DMSO) alone as 1. A. Aldosterone, 11-deoxycorticosterone, 11-deoxycortisol and progesterone with full-length lungfish MR with TAT3-luc. B. Aldosterone, 11-deoxycorticosterone, 11-deoxycortisol and progesterone with truncated lungfish MR (Domains CDE) with TAT3-luc. C. Cortisol, corticosterone, dexamethasone and triamcinolone with full-length lungfish MR with TAT3-luc. D. Cortisol, corticosterone, dexamethasone and triamcinolone with truncated lungfish MR (Domains CDE) with TAT3-luc.

**Table 1. EC50 values for steroid activation of full-length and truncated lungfish MR with the TAT3 promoter.**

|  | Aldosterone EC50 | 11-deoxycorticosterone EC50 | 11-deoxycortisol EC50 | Progesterone EC50 |
|---|---|---|---|---|
| **MR-full length** | 0.04 nM | 0.04 nM | 0.17 nM | 0.03 nM |
| **95% confidence interval** | 0.02-0.07 nM | 0.02-0.09 nM | 0.1-0.3 nM | 0.02-0.06 nM |
| **MR-CDE** | 0.24 nM | 0.13 nM | 0.27 nM | 0.044 nM |
| **95% confidence interval** | 0.17-0.35 nM | 0.08-0.2 nM | 0.14-0.53 nM | 0.026-0.076 nM |

|  | Corticosterone EC50 | Cortisol EC50 | Triamcinolone EC50 | Dexamethasone EC50 |
|---|---|---|---|---|
| **MR-full length** | 23.1 nM | 66.1 nM | 1.3 nM | 4.7 nM |
| **95% confidence interval** | 11.3-47.1 nM | 44.5-98.4 nM | 0.9-1.9 nM | 3.3-6.9 nM |
| **MR-CDE** | 85.5 nM | 86.7 nM | 2.4 nM | 7.7 nM |
| **95% confidence interval** | 60.0-121.8 nM | 74.3-101.2 nM | 2.1-2.8 nM | 6.6-9.1 nM |

**Transcriptional activation of full-length and truncated human MR and full-length and truncated elephant shark MR by corticosteroids and progestins.**

To gain an evolutionary perspective on activation of lungfish MR by steroids, we screened a panel of steroids, at 10 nM, for transcriptional activation of full-length human and elephant shark MRs and truncated human and elephant shark MR-CDEs using two reporters: MMTV-luc and TAT3-luc.

**Comparison of human MR and lungfish MR.**

Overall, compared to lungfish MR, fold activation of human MR was significantly higher for aldosterone and other corticosteroids. For example, compared to 2-fold activation by aldosterone of full-length lungfish MR with the MMTV promoter (Figure 3A), activation of full-length human MR by aldosterone was about 70-fold with the MMTV promoter (Figure 6A). Although fold-activation by steroids for truncated human MR (Figure 6B) decreased compared to full-length human MR (Figure 6A), activation by aldosterone and other corticosteroids of truncated human MR with the MMTV promoter (Figure 6B) was about 7-fold higher than for truncated lungfish MR (Figure 3B).

Unlike for lungfish MR, deletion of the NTD in human MR resulted in a loss of activation by aldosterone and other corticosteroids for human MR-CDE with both promoters (Figure 6A-D), consistent with the presence of two activation function domains in the NTD (Figure 2) [49–52]. The relative loss of activation of human MR was greater with the MMTV promoter than with the TAT3 promoter. For example, at 10 nM aldosterone, activation of full-length human MR with the MMTV reporter was 70-fold (Figure 6A), which decreased to 14-fold for human MR-CDE (Figure 6B). In contrast, at 10 nM aldosterone, fold-activation of human MR-CDE with the TAT3 promoter was about 75% of activity for full-length human MR (Figure 6C, D). However, 11-deoxycorticosterone and 11-deoxycortisol lost substantial activity for human MR-CDE with the MMTV and TAT3 promoters (Figure 6D).

There also was higher fold-activation by aldosterone of full-length and truncated human MR with the TAT3 promoter (Figure 6C, D) compared to full-length and truncated lungfish MR (Figure 3C, D). Aldosterone activation of full-length human MR with the TAT3 promoter (Figure 6C) was about 45-fold higher than that for full-length lungfish MR with the TAT3 promoter (Figure 3C). Aldosterone activation of human MR-CDE with the TAT3 promoter (Figure 6D) was about 15-fold higher than that for lungfish MR-CDE (Figure 3D).

The relative activation by aldosterone and dexamethasone of human MR and lungfish MR was reversed. Aldosterone was more active than dexamethasone in stimulating transcription by full-length human MR and human MR-CDE with the TAT3 promoter (Figure 6C, D). In contrast, for lungfish MR dexamethasone was more active than aldosterone for full-length lungfish MR and lungfish MR-CDE with the TAT3 promoter (Figure 3C, D).

**Comparison of elephant shark MR and lungfish MR.**

Activation by corticosteroids and progesterone of elephant shark MR with the MMTV promoter has some similarities with their activation of lungfish MR. Like lungfish MR, corticosteroids have a similar activation of about 10-fold for full-length and truncated elephant shark MR, with little difference in potency among the corticosteroids. However, unlike lungfish MR, aldosterone is stronger than dexamethasone in activating full-length and truncated elephant shark MR with the MMTV promoter (Figure 6E, F).

At a 10 nM steroid concentration, aldosterone and other corticosteroids activated full-length elephant shark MR with the TAT3 promoter by 9 to 12-fold (Figure 6G), which was similar to activation with the MMTV promoter (Figure 6E). Activation of full-length elephant shark MR by progesterone was about 5-fold with the TAT3 and MMTV promoters (Figure 6E, G). Aldosterone was about 2-fold more active than dexamethasone.

However, deletion of the NTD from elephant shark MR resulted in a significant increase is activation by steroids in the presence of the TAT3 promoter (Figure 6H). Thus, truncated elephant shark MR with the TAT3 promoter was activated from 300 to 350-fold by aldosterone and other corticosteroids and about 200-fold by progesterone and dexamethasone (Figure 6H), indicating that like lungfish MR, the NTD in elephant shark inhibits activation by corticosteroids. However, unlike lungfish MR, compared to aldosterone, dexamethasone was less active for full-length and truncated elephant shark MR with the MMTV promoter and truncated elephant shark MR with the TAT3 promoter.

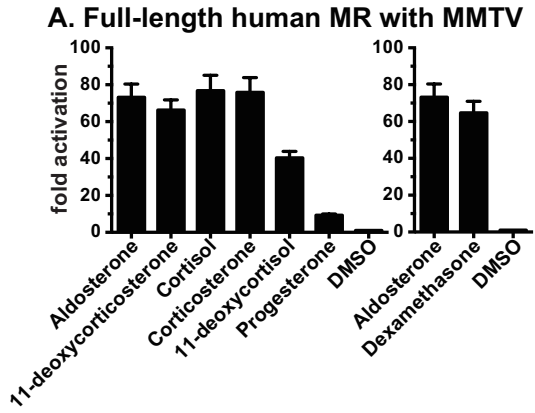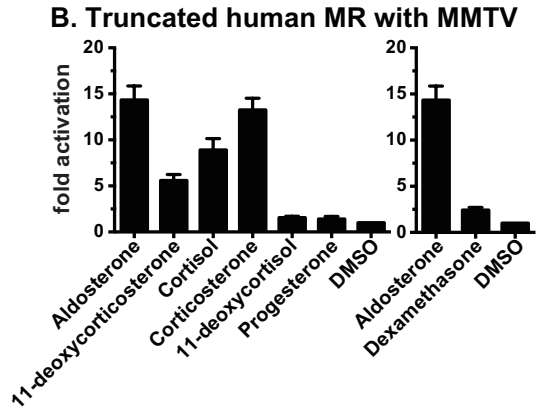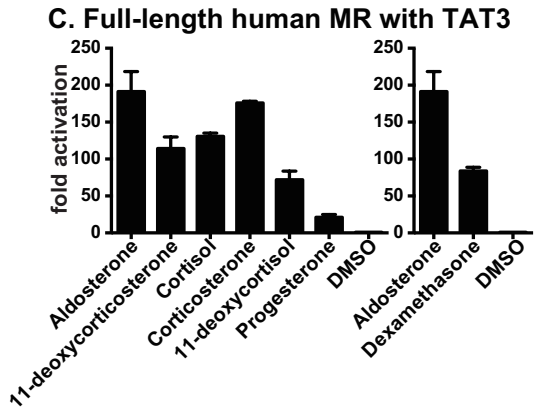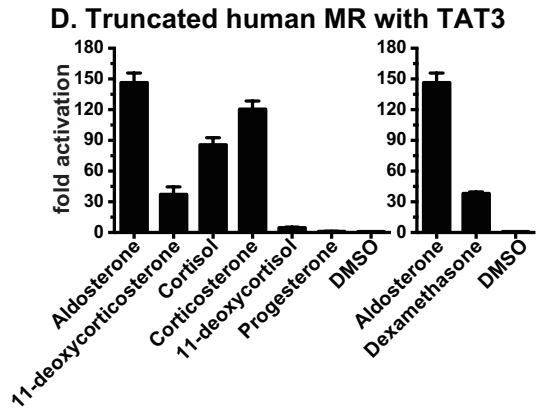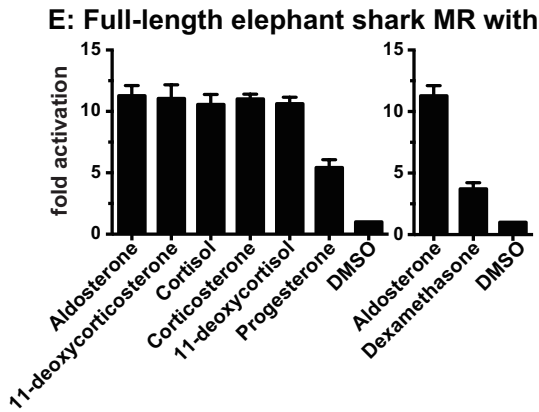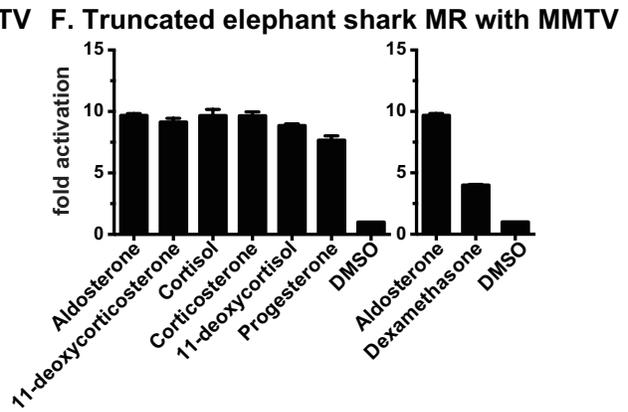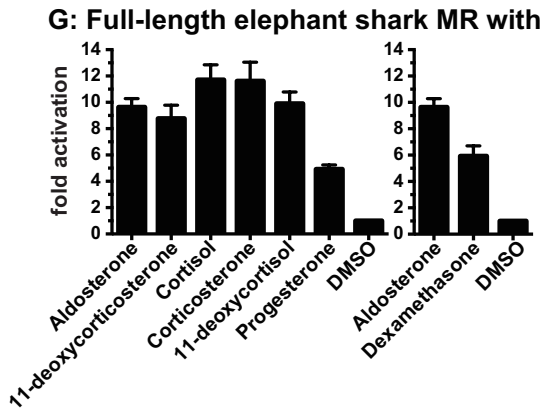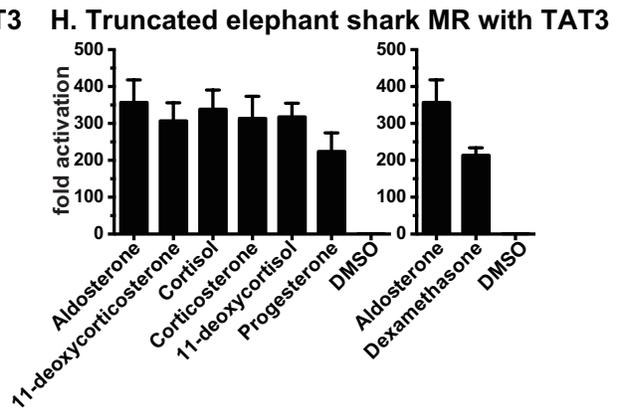

**Figure 6. Ligand specificity of full-length and truncated human MR and elephant shark MR.** Plasmids for full-length human and elephant shark MR or truncated human and elephant shark MR (MR-CDE) were expressed in HEK293 cells with an MMTV-luciferase reporter or a TAT3-luciferase reporter. Transfected cells were treated with either 10 nM aldosterone, cortisol, 11-deoxycortisol, corticosterone, 11-deoxycorticosterone, progesterone, dexamethasone or vehicle alone (DMSO). Results are expressed as means ± SEM, n=3. Y-axis indicates fold-activation compared to the activity of control vector with vehicle alone as 1. A. Full-length human MR with MMTV-luciferase. B. Truncated human MR (MR-CDE) with MMTV-luciferase. C. Full-length elephant shark MR with TAT3-luciferase. D. Truncated elephant shark MR (MR-CDE) with TAT3-luciferase.

**Does lungfish contain a separate GR gene?**

We used sequences of human GR, coelacanth GR and elephant shark GR as probes in a BLAST search of GenBank and did not retrieve a lungfish GR sequence. The absence of a lungfish GR ortholog coupled with the strong response of lungfish MR to dexamethasone and triamcinolone (Figure 5) leads us to propose that lungfish MR also functions as a GR.

**Discussion**

Dobzhansky's aphorism "Nothing in Biology Makes Sense Except in the Light of Evolution" [62] explains the importance of the evolution of aldosterone in lungfish because aldosterone activation of the kidney MR in terrestrial vertebrates regulates sodium, potassium and water transport, which is critical in maintaining internal electrolyte homeostasis in terrestrial vertebrates [30,34,63–65] an activity that was important in the transition from water to land. Here we report that aldosterone, 11-deoxycorticosterone, and progesterone have EC50s below 1 nM for lungfish MR (Table 1), which makes these steroids potential physiological ligands for lungfish MR. Another potential physiological steroid is 11-deoxycortisol, which is a steroid for the CR in Atlantic sea lamprey [42,43]. 11-deoxycortisol has EC50 of 0.17 nM for full-length lungfish MR (Table 1).

A functional advantage of 11-deoxycorticosterone, 11-deoxycortisol and progesterone as ligands for the MR is that they lack an 11β-hydroxyl group, and thus, like aldosterone, they are inert to 11β-hydroxysteroid dehydrogenase-type 2, unlike cortisol and corticosterone [66–69]. Indeed, this inertness to 11β-hydroxysteroid dehydrogenase-type 2 and the low EC50s of these steroids for lungfish MR suggests that more than one corticosteroid and progesterone [39] may be physiological mineralocorticoids.

Like ray-finned fish MRs [24,37,38,60] and elephant shark MR [24,39], lungfish MR is activated by spironolactone (Figure 4), and, as reported here, by eplerenone [47,48,61].

We also find important differences between the response of lungfish MR and human MR to aldosterone, 11-deoxycorticosterone, 11-deoxycortisol and progesterone, indicating that further selectivity for aldosterone in human MR occurred during the evolution of terrestrial vertebrates [13,15,37,40,52,69,70].

An unexpected difference between lungfish MR and human MR is the substantial increase in fold-activation by steroids of lungfish MR after deletion of the NTD, in contrast to human MR in which the NTD contains an activation function domain (Figure 2) [49–52]. Deletion of the NTD in elephant shark MR also resulted in a substantial increase in fold-activation by corticosteroids and progesterone using the TAT3 promoter (Figure 6), but not for the MMTV promoter [17]. These data with lungfish MR and elephant shark MR suggest that early in the evolution of the MR there was an allosteric interaction between the LBD and NTD [71,72] that repressed steroid activation of the MR, and that the activation function in the NTD as found in human MR [49–52] evolved later in terrestrial vertebrates, along with changes in steroid specificity, such loss of MR activation by progesterone [37,39,70]. The different responses of full-length and truncated lungfish MR, human MR and elephant shark MR with the MMTV and TAT3 promoters indicate that the NTD and the promoter are important regulators of steroid activation of these MRs. Corticosteroid activation of these MRs in the presence of other promoters merits investigation.

The stronger response of lungfish MR to dexamethasone compared to aldosterone and the absence a lungfish GR ortholog sequence are puzzling. At a 10 nM concentration, fold-activation by dexamethasone and triamcinolone is substantially higher than that of cortisol, corticosterone, as well as aldosterone, 11-deoxycorticosterone, 11-deoxycortisol and progesterone for lungfish MR (Figure 5). One explanation is that lungfish MR also has a GR function.

## Materials and Methods

### Chemical reagents

Aldosterone, cortisol, corticosterone, 11-deoxycorticosterone, 11-deoxycortisol and progesterone, spironolactone and eplerenone were purchased from Sigma-Aldrich. For reporter gene assays, all hormones were dissolved in dimethyl-sulfoxide (DMSO); the final DMSO concentration in the culture medium did not exceed 0.1%.

### Animal

A slender spotted African lungfish, *Protopterus dolloi,* was purchased from a local commercial supplier. Lungfish were anesthetized in freshwater containing 0.02% ethyl 3-aminobenzoate methane-sulfonate from Sigma-Aldrich, and tissue samples were quickly dissected and frozen in liquid nitrogen. Animal handling procedures conformed to the guidelines set forth by the Institutional Animal Care and Use Committee at the University of Tokyo.

**Molecular cloning of lungfish mineralocorticoid receptor**

Two conserved amino acid regions, GCHYGV and LYFAPD of vertebrate MRs were selected and degenerate oligonucleotides were used as primers for PCR. First-strand cDNA was synthesized from 2 μg of total RNA isolated from the liver after amplification, and an additional primer set (CKVFFK and LYFAPD) was used for the second PCR. The amplified DNA fragments were subcloned with TA-cloning plasmid pGEM-T Easy vector, sequenced using a BigDye terminator Cycle Sequencing-kit with T7 and SP6 primers, and analyzed on the 3130 Genetic Analyzer (Applied Biosystems). The 5'- and 3'-ends of the mineralocorticoid receptor cDNAs were amplified by rapid amplification of the cDNA end (RACE) using a SMART RACE cDNA Amplification kit. Genbank accessions for this lungfish MR are: Nucleotide ID: LC630795 and Protein ID: BCV19931.

**Construction of plasmid vectors**

The full-length and truncated MRs were amplified by PCR with KOD DNA polymerase. The PCR products were gel-purified and ligated into pcDNA3.1 vector (Invitrogen). The truncated MR proteins were designed to possess methionine and valine residues at the N-terminus and contain a DNA-binding domain, a hinge-region, and a ligand-binding domain. The truncated MRs were amplified by PCR with KOD DNA polymerase by using the following primers: lungfish MR forward primer (5'-CAAGCTTACCATGGTGTGTCTGGTGTGTGGTGACGAAG-3' containing *Hind*III site) and lungfish MR reverse primer (5'-CCTACTTCCTGTGAAAGTACAATGAC -3' containing stop codon), human MR forward primer (5'-CGGATCCACCATGGTGTGTTTGGTGTGTGGGGATGAG-3' containing *Bam*HI site) and human MR reverse primer (5'-CTCACTTCCGGTGGAAGTAGAGCGGC -3' containing stop codon). The amplified DNA fragments were subcloned with TA-cloning plasmid pGEM-T Easy vector and sequenced, and then subcloned into pcDNA 3.1 vector by using *Hind*III-*Not*I sites for lungfish MR truncated form or *Bam*HI-*Not*I sites for human MR truncated form. Mouse mammary tumor virus-long terminal repeat (MMTV-LTR) was amplified from pMSG vector by PCR, and inserted into pGL3-basic vector containing the *Photinus pyralis* lucifease gene. 3X-Tyrosine Amino Transferase (TAT3) promoter containing reporter vector named pGL4.23-TAT3-Luc was constructed as described previously [53]. All cloned DNA sequences were verified by sequencing.

**Transactivation assay and statistical methods**

Transfection and reporter assays were carried out in HEK293 cells, as described previously [40,73]. All experiments were performed in triplicate. The values shown are mean ± SEM from three separate experiments, and dose-response data, which were used to calculate the half maximal response (EC50) for each steroid, were analyzed using GraphPad Prism.

## DECLARATION OF INTEREST

We have no conflict of interest.

## ACKNOWLEDGEMENT

We thank Peter Fuller, Ron de Kloet, Patrick Prunet and Bernard Rossier for constructive comments about an earlier version of this manuscript.

## AUTHOR CONTRIBUTIONS

Y.K., S.O., and M.E.B. carried out the research and analyzed data. S.H. aided in the collection of animals. X.L. constructed plasmid DNAs used in this study. Y.K. and M.E.B. conceived and designed the experiments. Y.K. and M.E.B. wrote the paper. All authors gave final approval for publication.

## FUNDING

Y.K. was supported in part by Grants-in-Aid for Scientific Research [19K067309] from the Ministry of Education, Culture, Sports, Science and Technology of Japan, and Takeda Science Foundation. M.E.B. was supported by Research fund #3096.
## REFERENCES

1. Baker ME, Nelson DR, Studer RA. Origin of the response to adrenal and sex steroids: Roles of promiscuity and co-evolution of enzymes and steroid receptors. J Steroid Biochem Mol Biol. 2015;151:12-24. doi:10.1016/j.jsbmb.2014.10.020.

2. Evans RM. The steroid and thyroid hormone receptor superfamily. Science. 1988;240(4854):889-895. doi:10.1126/science.3283939.

3. Bridgham JT, Eick GN, Larroux C, et al. Protein evolution by molecular tinkering: diversification of the nuclear receptor superfamily from a ligand-dependent ancestor. PLoS Biol. 2010;8(10):e1000497. Published 2010 Oct 5. doi:10.1371/journal.pbio.1000497.

4. Baker ME, Funder JW, Kattoula SR. Evolution of hormone selectivity in glucocorticoid and mineralocorticoid receptors [published correction appears in J Steroid Biochem Mol Biol. 2014 Jan;139:104]. J Steroid Biochem Mol Biol. 2013;137:57-70. doi:10.1016/j.jsbmb.2013.07.009.

5. Lifton RP, Gharavi AG, Geller DS. Molecular mechanisms of human hypertension. Cell. 2001;104(4):545-556. doi:10.1016/s0092-8674(01)00241-0.
20